\documentclass[prb,twocolumn,superscriptaddress,nobibnotes,amsmath,amssymb]{revtex4-1}
\usepackage{natbib}
\usepackage{textcomp}
\usepackage{float}
\usepackage[utf8]{inputenc}
\usepackage{braket}
\usepackage{array}
\usepackage{graphicx}         
\usepackage{bm}               
\usepackage{color}
\definecolor{darkblue}{rgb}{0.0,0.0,0.3}
\definecolor{goodblue}{rgb}{0.0,0.0,0.6}
\usepackage{float}
\usepackage{hyperref}
\usepackage{upgreek}
\usepackage{gensymb}
\hypersetup{
	colorlinks   = true, 
	urlcolor     = darkblue, 
	linkcolor    = darkblue, 
	citecolor   = darkblue 
}

\makeatletter
\def\vhrulefill#1{\leavevmode\leaders\hrule\@height#1\hfill \kern\z@}
\makeatother

\begin{document}
\thispagestyle{empty}
\onecolumngrid
\begin{center}
	\textbf{\large Universal Platform for Scalable Semiconductor-Superconductor Nanowire Networks}
\end{center}
\begin{center}
\normalsize	{			
			Jason Jung,$^{1,*}$ Roy L.M. Op het Veld,$^{1,*}$ Rik Benoist,$^1$ Orson A.H. van der Molen,$^1$ Carlo Manders,$^2$  Marcel A. Verheijen,$^{1,2}$ Erik P.A.M. Bakkers$^{1,\dagger}$			
			}
\smallskip
\small

\emph{$^\mathit{1}$Department of Applied Physics, Eindhoven University of Technology, Groene Loper 19, 5612AP Eindhoven, The Netherlands}

\emph{$^\mathit{2}$Eurofins Materials Science Netherlands BV, High Tech Campus 11, 5656 AE Eindhoven, The Netherlands}

\vspace*{-0.1cm}
\end{center}

\begin{abstract}
	Semiconductor-superconductor hybrids are commonly used in research on topological quantum computation. Traditionally, top-down approaches involving dry or wet etching are used to define the device geometry. These often aggressive processes risk causing damage to material surfaces, giving rise to scattering sites particularly problematic for quantum applications. Here, we propose a method that maintains the flexibility and scalability of selective area grown nanowire networks while omitting the necessity of etching to create hybrid segments. Instead, it takes advantage of directional growth methods and uses bottom-up grown InP structures as shadowing objects to obtain selective metal deposition. The ability to lithographically define the position and area of these objects, and to grow a predefined height, ensures precise control of the shadowed region. We demonstrate the approach by growing InSb nanowire networks with well-defined Al and Pb islands. Cross-section cuts of the nanowires reveal a sharp, oxide-free interface between semiconductor and superconductor. By growing InP structures on both sides of in-plane nanowires, a combination of Pt and Pb can independently be shadow deposited, enabling a scalable and reproducible in-situ device fabrication. The semiconductor-superconductor nanostructures resulting from this approach are at the forefront of material development for Majorana based experiments.
\end{abstract}

\maketitle
\twocolumngrid
\normalsize

Material science is a major impetus for developments in topological quantum computation. This active research field aims to use non-abelian anyons to provide a basis for naturally fault-tolerant quantum computing.\cite{RN707, RN1} A major step towards the first practical realization was the prediction that these quasiparticles can form in a one-dimensional semiconductor with strong spin-orbit Rashba interaction, coupled to a conventional \mbox{s-wave} superconductor.\cite{RN4, RN3, RN5} Several proposals have emerged on how these heterostructures can be used to build a fully functional topological qubit, most of which are based on branched nanowire networks with topological segments.\cite{RN6, RN708} This sets the fabrication of these networks as one of the major challenges for material scientists in the field.

In conventional semiconductor microfabrication, top-down approaches involving dry or wet etching are used to define device geometries.\cite{RN8} These processes are often detrimental to the semiconductor surface, giving rise to disorder on which the carriers can scatter. This is particularly problematic for quantum applications, where these scattering sites negatively impact carrier mobility and coherence length. The preferred alternative is the bottom-up approach, in which the device structure is built from its individual atoms self-arranging into the desired pattern. An archetype of bottom-up synthesis is the growth of out-of-plane nanowires. They show ballistic transport over several microns\cite{RN10, RN9} and have been used in many successful Majorana based experiments.\cite{RN12, RN13, RN14, RN11, RN15, RN16} Shadow deposition of superconductors has furthermore successfully been used, to avoid etching damage and the need of developing etching procedures.\cite{RN730, RN17, RN20, RN19} The downside of these 

\vspace{.1cm}
\noindent \vhrulefill{0.25pt} \hspace*{7.2cm}
\vspace{.4cm}

\footnotesize
\noindent $^*$ These authors contributed equally to this work.\\
$^{\dagger}$ Correspondence to E.P.A.M. (\href{mailto:e.p.a.m.bakkers@tue.nl}{e.p.a.m.bakkers@tue.nl}).

\normalsize
\noindent out-of-plane synthesized wires is the limited geometric flexibility in creating networks. 

Recently, a more scalable approach was developed based on the selective-area growth technique (SAG). In this method, an amorphous mask is deposited on the growth substrate into which the network pattern is etched. The right growth conditions confine the subsequent nanowire synthesis to the mask openings, allowing for scalable and highly flexible growth of complex in-plane selective area networks.\cite{RN23, RN37, RN322} But the difficulty of creating semiconductor-superconductor segments remains. Until now, Al was subsequently evaporated and selectively wet etched, a method frequently used on out-of-plane nanowires.\cite{RN322, RN24} Experiments show however, that even highly selective etchants inflict damage on the semiconductor surface, lowering the device performance.\cite{RN730, RN19}

\begin{figure*}[ht]
	\centering
	\includegraphics[width=1\textwidth]{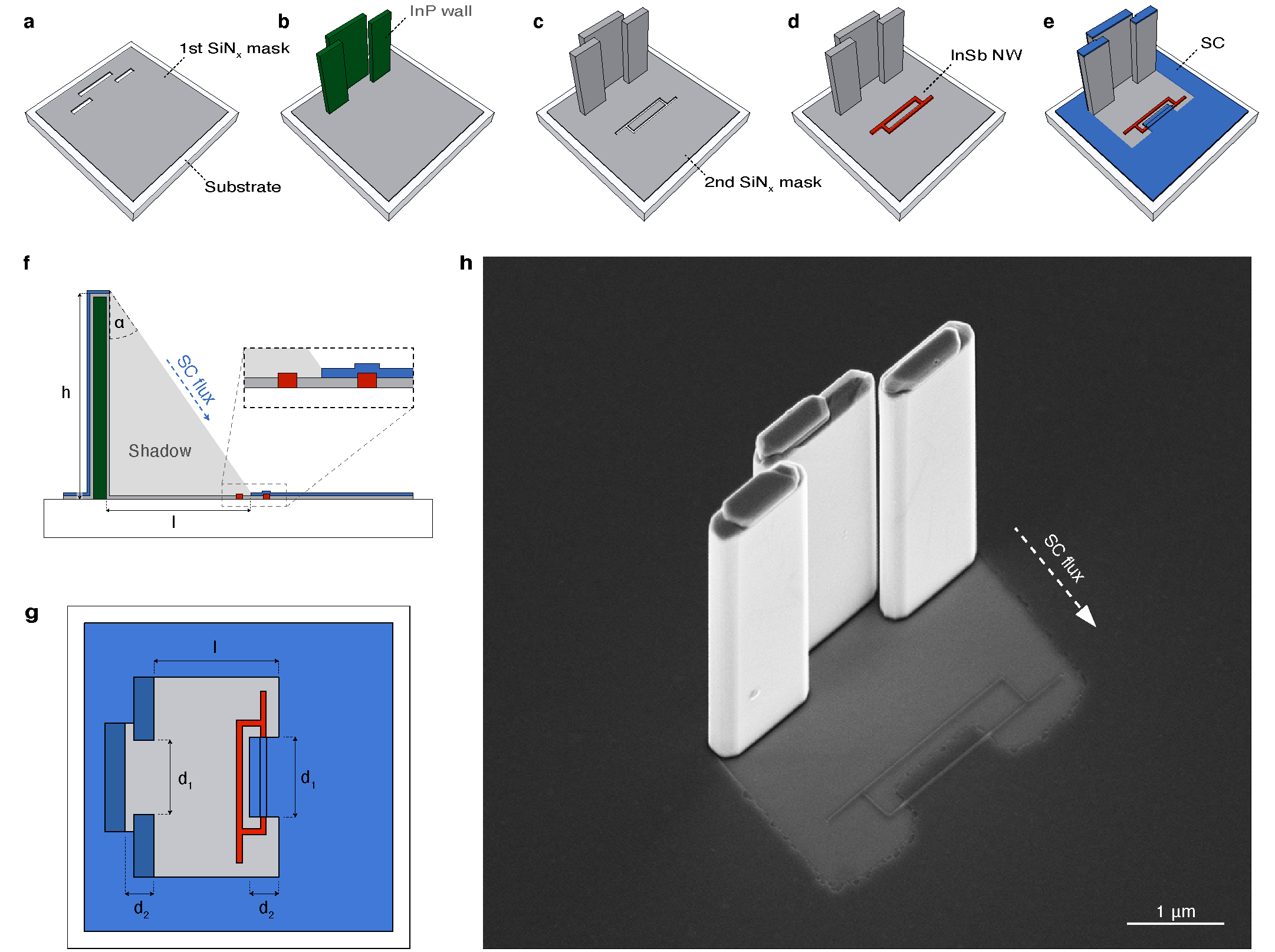}
	\caption{\textbf{Shadow deposit networks using InP walls.} \textbf{a-e,} Illustration of the growth principle. Lithographically defined openings are etched into a Si$_{\text{x}}$N$_{\text{y}}$ mask~(grey) on an InP~(111)A substrate~(white), followed by selective-area growth of InP walls~(green). The same approach is used to selectively grow the InSb~(red) network after depositing and etching a second mask. The InP walls act as shadowing objects during the directional superconductor~(blue) deposition to allow its selective growth on the nanowire network. \textbf{f,} A cross-section of the design shows the dependency of the shadow length~$l$ on wall height~$h$ and deposition angle~$\alpha$. The inset shows a close up of the partially covered network. \textbf{g,} Top-view schematic, indicating the design flexibility of the approach due to the precise control over position and size of walls and network. \textbf{h,} Tilt-view SEM image of an in-plane InSb nanowire network with a superconducting Al island created through shadow deposition using InP walls. The arrow indicates the direction of the SC flux during deposition. The Al is deposited at $\alpha = 25\degree$.}
\end{figure*}

In this article we present the integration of shadow deposition with SAG, allowing for a reproducible and scalable bottom-up approach. We demonstrate the growth of branched, in-plane InSb nanowire networks and full control over number, size, and position of its superconducting segments. Furthermore, our method is suitable for the directional deposition of a broad range of materials. To show this, we epitaxially grow thin films of Al, the most established superconductor choice within the field, as well as Pb. The latter possesses the highest bulk critical field and temperature of all elemental type-I superconductors, and has shown high promise in recent experiments.\cite{RN748} Taking the same approach to shadow networks from multiple sides allows for the combination of several shadow deposited materials. We demonstrate this through various examples, including a full device consisting of an InSb nanowire, Pb islands, and shadow deposited Pt leads. Altogether, this method provides an ideal basis for the next wave of Majorana based experiments.

Figure~1 illustrates the principle of the technique. As an example, we show the fabrication of an InSb Aharonov-Bohm interferometer with an embedded Al island. This type of device has raised significant interest due to its predicted non-local phase-coherent electron transfer.\cite{RN26, RN27, RN28} The schematics in Figure 1a-e give an overview of the fabrication steps. The starting point is a semi-insulating InP~(111)A substrate covered by a 20~nm amorphous Si$_{\text{x}}$N$_{\text{y}}$ mask. Lithographically defined openings are dry etched into the mask along the $\langle10\bar{1}\rangle$ and $\langle11\bar{2}\rangle$ crystal directions, allowing access to the underlying crystalline substrate. The subsequent metalorganic vapor-phase epitaxy (MOVPE) forms InP nanostructures, herein referred to as walls.\cite{RN29} Confining the mask openings to these crystal directions allows for high control over the lateral dimensions of the walls (see Supplementary Information 1). A second Si$_{\text{x}}$N$_{\text{y}}$ mask is deposited, and the same method is used to first etch the mask opening and then selectively grow the InSb nanowire network using MOVPE.\cite{RN37, RN322} The initially grown InP walls act as shadowing objects during the subsequent superconductor deposition and allow for selective growth on the InSb network. 

\begin{figure*}[ht]
	\centering
	\includegraphics[width=1\textwidth]{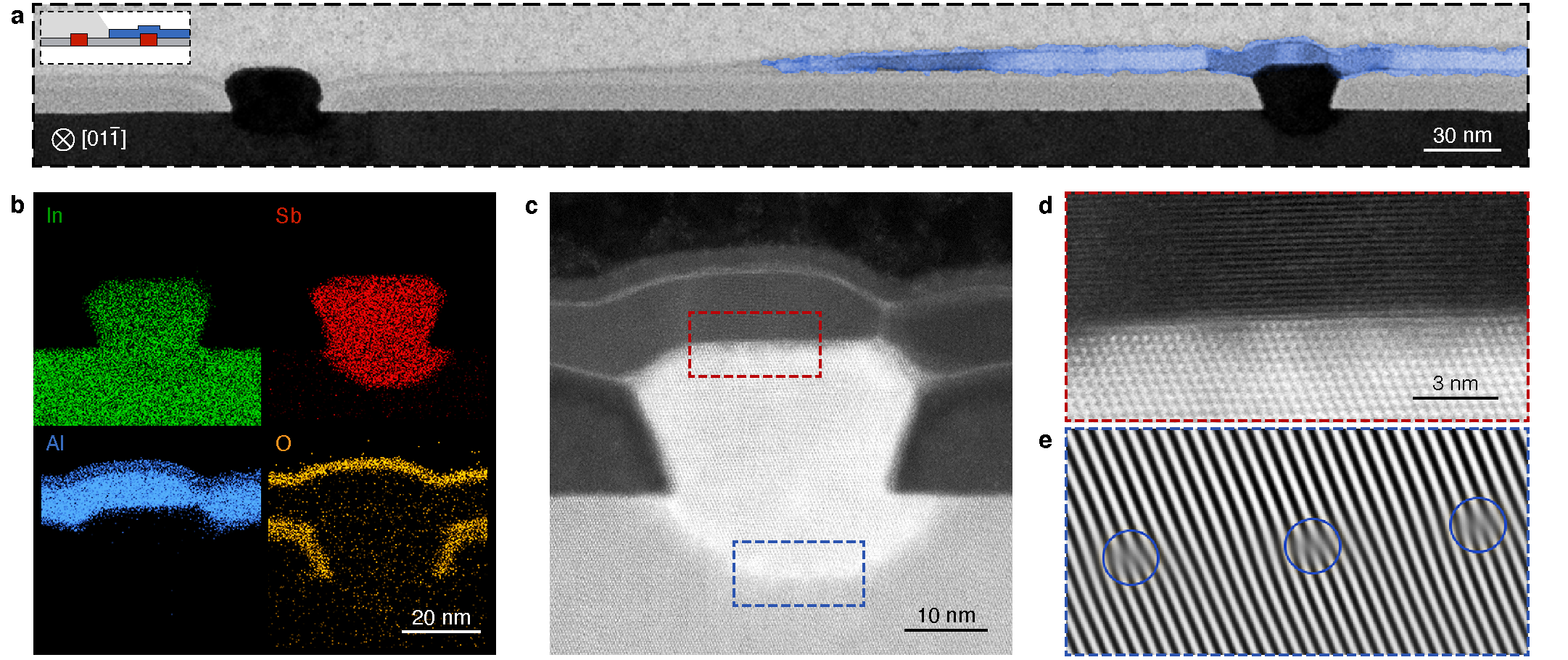}
	\caption{\textbf{Partial shadowing.} \textbf{a,} Bright Field scanning transmission electron micrograph from a cross-section cut of the device shown in Figure~1. The partial shadowing of the network is visible with only the right wire being covered by the Al superconductor. This is confirmed by an overlayed EDX elemental mapping of Al (blue). The inset refers back to the schematic of Figure~1f. \textbf{b,} EDX elemental mapping of the covered nanowire with In (green), Sb (red), Al (blue), and oxygen (orange), showing that the Al layer is continuous from the nanowire to the mask. \textbf{c,} HAADF STEM image of the covered nanowire. \mbox{\textbf{d,} High} resolution HAADF STEM of the region marked with a red rectangle in panel \textbf{d} depicting a uniform, oxide free InSb/Al interface. \textbf{e,} IFFT of the region marked with a blue rectangle in panel \textbf{d} reveals the atomic columns at the InP substrate to InSb nanowire interface. The lattice mismatch between InP and InSb is compensated by misfit dislocations directly at the interface (blue circles).}
\end{figure*}

The angle $\alpha$ relative to the wafer normal, under which the superconductor is deposited, and the height h of the InP nanostructure determine the length $l=h \sin{\alpha}$ of the shadow, as depicted in Figure~1f. The ability to lithographically define the lateral dimensions and positions of the InP nanostructures and the InSb network, combined with the control of the wall height during MOVPE growth, gives precise control over the shadowed region. This is indicated in Figure~1g, where the distances $d_1$ and~$d_2$ between the nanostructures define the length and position of the metal island, ensuring that the shadow falls between the two arms of the interferometer. A scanning electron microscopy (SEM) image of a device created using this technique is shown in Figure~1h. The result is a single crystalline InSb nanowire interferometer with a well-defined $\sim1.2~\upmu$m long Al island on one arm. In addition to this, many other networks were fabricated to demonstrate the flexibility, accuracy and scalability of the approach (see Supplementary Information 2 and 3). This includes the basis of a proposed four-qubit device consisting of twelve superconducting segments connected into a branched network.\cite{RN708}

A transmission electron microscopy (TEM) image of the cross-section of the Aharonov-Bohm interferometer is shown in Figure~2a. The partial shadowing of the network is visible, with a $\sim8$~nm thick Al film covering only the right nanowire. This is confirmed through an energy-dispersive x-ray spectroscopy (EDX) elemental mapping of the same region. A more detailed EDX mapping of the right wire in Figure~2b shows an uninterrupted Al film from the top of the nanowire to the Si$_{\text{x}}$N$_{\text{y}}$ mask. This allows for contacting the superconducting segment of the network without fabricating directly on the nanowire itself, thereby preventing subsequent damage. Other designs specifically require disconnecting the metal layer on the network from the layer on the mask to electrically isolate superconducting segments within the network. This can be achieved by growing the nanowire itself to a height sufficient to enable self-shadowing (see Supplementary Information 4).

\begin{figure*}[ht]
	\centering
	\includegraphics[width=1\textwidth]{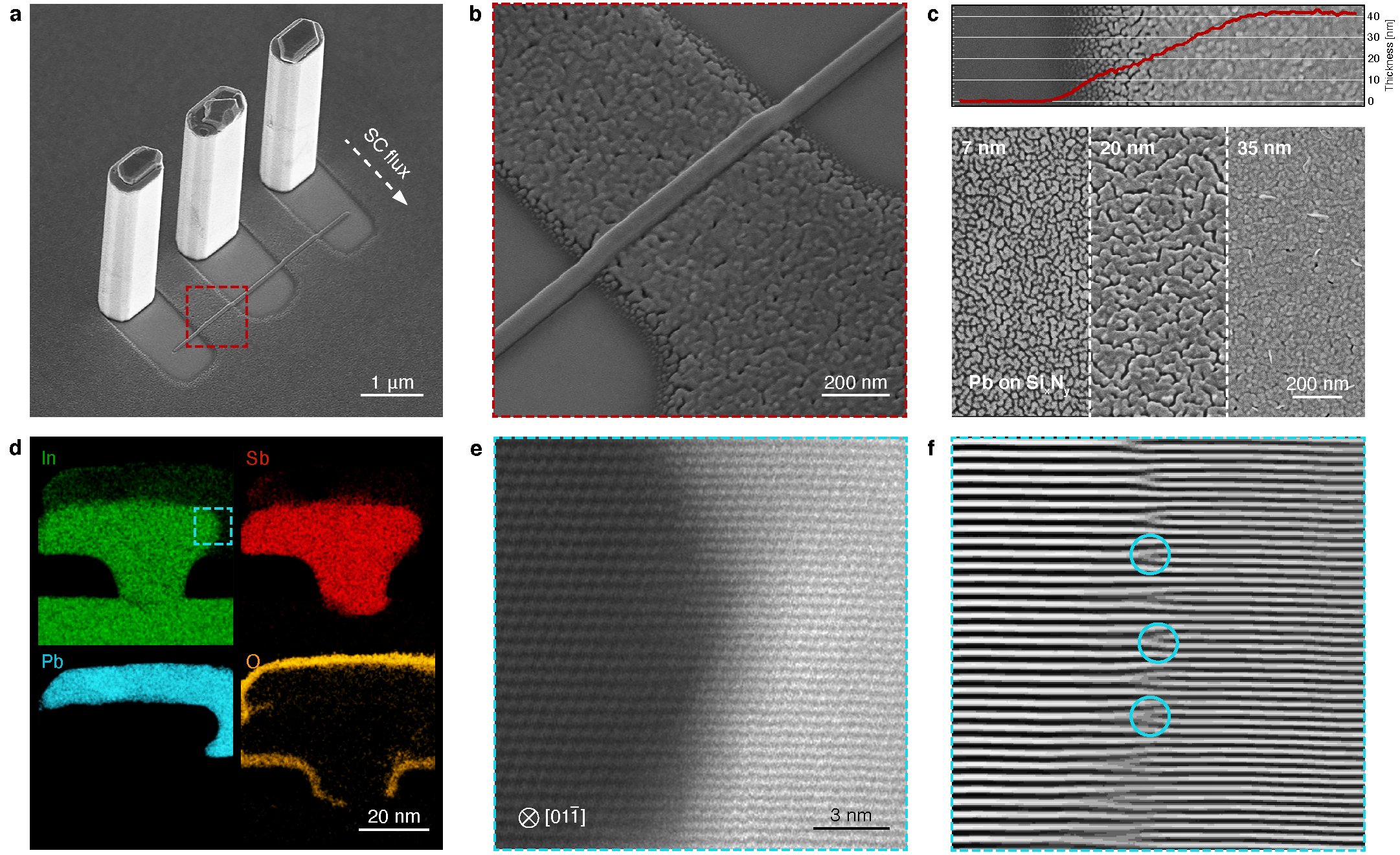}
	\caption{\textbf{Epitaxial Pb on InSb.} \textbf{a,} Tilt-view SEM image of an InSb nanowire with two shadow deposited $\sim22$~nm thick Pb islands. The arrow indicates the direction of the superconductor flux during deposition. The Pb is deposited at $\alpha = 25\degree$. \textbf{b,} Close up of the left island showing a smooth, closed Pb film on the crystalline InSb nanowire and a rough poly-crystalline topography on the amorphous Si$_{\text{x}}$N$_{\text{y}}$ mask. \textbf{c,} Layer morphology of Pb on the Si$_{\text{x}}$N$_{\text{y}}$ mask. With increasing thickness, the film transitions from separate islands, to a continuous rugged layer, to eventually forming a closed film. \textbf{d,} EDX elemental mapping of a cross-section imaged along the $[01\bar{1}]$ zone axis with In (green), Sb (red), Pb (cyan), and oxygen (orange). \textbf{e,} HAADF scanning TEM image of the region marked with a blue rectangle in panel \textbf{d} showing the oxide free InSb/Pb interface. \textbf{f,} IFFT of the same region reveals the 3:4 ratio of the epitaxially aligned $(111)_{\text{InSb}}$ and $(111)_{\text{Pb}}$ planes with a residual lattice mismatch of 1.7\%. Three exemplary edge dislocations are indicated by circles (cyan).}
\end{figure*}

A high-angle annular dark-field (HAADF) scanning TEM image of the covered nanowire is shown in Figure~2c. A gentle removal of the native InSb oxide using a directional flow of atomic hydrogen radicals prior to the superconductor deposition ensures the clean semiconductor-superconductor interface\cite{RN30}, as seen in Figure~2d. This the first reported selective-area growth of InSb nanowires on InP~(111)A (see Supplementary Information 5). The substrate polarity is a crucial requirement for the selective-area homoepitaxy of InP but had not been used for InSb SAG prior. The growth behaviour shows many parallels to InP~(111)B with the 10.4\% lattice mismatch between substrate and nanowire compensated by misfit dislocations directly at the interface.\cite{RN322} These dislocations can be revealed through an inverse fast Fourier transformation (IFFT) of the TEM image, as depicted in Figure~2e. 

The presented technique is not limited to the use of Al. One particularly promising superconductor alternative is Pb, due to its expected large superconducting gap, and high critical field and critical temperature.\cite{RN748} The SEM micrographs in Figure~3a-b show an in-plane InSb nanowire with $\sim22$~nm thick shadow deposited Pb forming two superconducting islands. This design is based on an experiment suggested to investigate Majorana fusion rules.\cite{RN32, RN31} The film growth is governed by an interplay of thermodynamics and kinetics. It is therefore highly dependent on the nature of the underlying layer. This is exemplified by the Pb forming a smooth, closed film on the crystalline InSb nanowire, while exhibiting a still continuous but rougher poly-crystalline morphology on the amorphous mask, with crevices at the grain boundaries. The morphology is determined by the minimization of the overall excess energy. Since Pb adatoms are more strongly bound to each other than to the Si$_{\text{x}}$N$_{\text{y}}$ mask they follow the Volmer-Weber growth mode.\cite{RN33} In this mode, three-dimensional islands nucleate directly on the surface after reaching a critical number of adatoms. Longer deposition time increases the island size until they coalesce and percolate, eventually forming a closed film. Figure~3c demonstrates the formation of the closed film with increasing film thickness. Creating a continuous superconducting layer reaching from the nanowire to the mask can be crucial for certain experiments, as previously indicated in the context of the Aharonov-Bohm interferometer. A nanowire with a single connected superconducting island (see Supplementary Information 3), could for instance serve as a material basis for three-terminal nonlocal conductance measurements.\cite{RN34, RN35} An EDX elemental mapping of a cross-sectional lamella of a comparable deposition is shown in Figure~3d. It exhibits a continuous Pb layer with a sharp, oxide-free, and epitaxial interface to the nanowire, as visible in the HAADF scanning TEM image depicted in Figure~3e. The InSb and Pb are imaged along the [011] axes of their zincblende and face centred cubic structures.  The IFFT in Figure~3f reveals a 3:4 ratio of the $(111)_{\text{InSb}}$ and $(111)_{\text{Pb}}$ lattice planes. Considering the $\text{d}(111)_{\text{InSb}} : \text{d}(111)_{\text{Pb}} = 0.763$ ratio of the interplanar spacings, the epitaxially matched InSb/Pb interface is estimated to have a residual lattice mismatch of 1.7\%.\cite{RN373, RN36}

\begin{figure}[t]
	\centering
	\includegraphics[width=.48\textwidth]{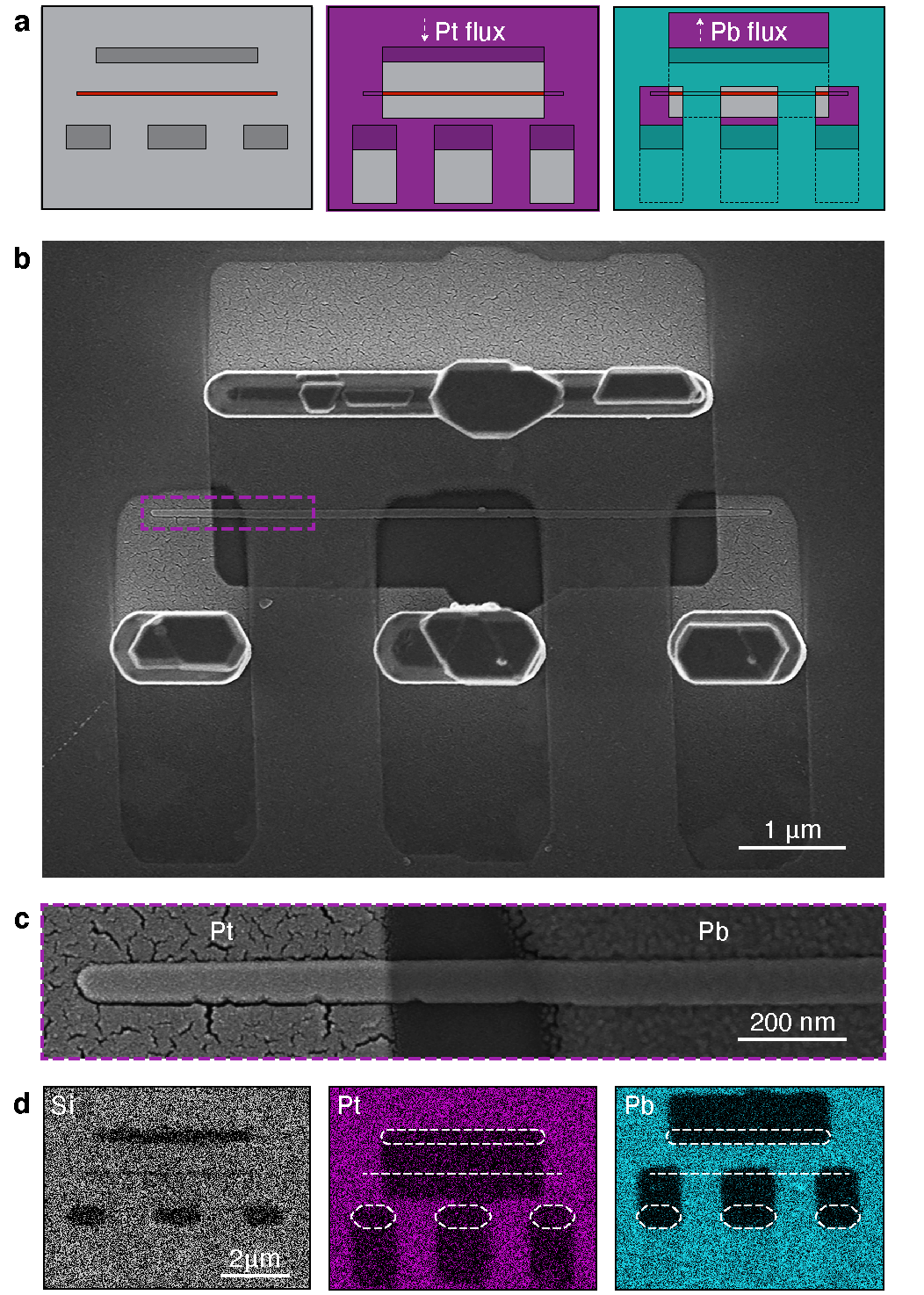}
	\caption{\textbf{Double-sided shadow deposition.} \textbf{a,} Schematic of double-sided shadow deposition. InP walls are placed on both sides of the nanowire. This allows for the consecutive shadow deposition of $\sim10$~nm Pt at $\alpha = 35\degree$ and $\sim22$~nm Pb at $\alpha = 25\degree$. \textbf{b,} Top-view SEM micrograph depicting the resulting InSb nanowire with two superconducting Pb islands and normal Pt segments on both ends of the wire. \textbf{c,} Close up of the normal-semiconductor-superconductor junction marked by a purple rectangle in panel \textbf{b}. It shows a sharp shadow with clear separation of the Pt and Pb covered segments by the bare InSb nanowire. \textbf{d,} SEM-EDX elemental mapping of the full device. The Si (grey) signal shows the etched wall and nanowire openings. The Pt (purple) and Pb (cyan) signals indicate the shadow deposited metals.}
\end{figure}

Finally, the shadowing approach is taken further by placing InP walls on both sides of the nanowire, allowing for the combination of multiple shadow deposited materials. An example to showcase the principle is sketched in Figure~4a. Here, the Pb island device introduced in Figure~3 is expanded by shadow depositing Pt on both ends of the nanowire. This creates a (double) junction between normal metal, semiconductor, and superconductor, which reduces the number of fabrication steps after the sensitive semiconductor-superconductor interface is created. An SEM micrograph of such a device is shown in Figure~4b-c. It exhibits sharp shadows with a clear separation between the Pt and Pb covered segments and the bare InSb nanowire. The dimensions of these segments can be accurately controlled through the position of the InP walls. Some applications require the bare InSb segments to be kept small in order to avoid the unintentional formation of quantum dots.\cite{RN13, RN38} To validate the accuracy of this approach, a Pt/Pb junction is created with a bare InSb segment of a lateral dimension as short as $\sim15$~nm (see Supplementary Information 6). The EDX analysis of the device is shown in Figure~4d, with the Si signal indicating the etched mask openings for walls and nanowire. The detected Pt and Pb signals reveal the deposition shadows and their relative position to the nanowire. The same approach is used to create a Pt/Al junction (see Supplementary Information 7). Next to combining normal metals with superconductors, this approach can also be used to combine two different superconductors for sensitive tunnel probing of the density of states (see Supplementary Information 8).\cite{RN40, RN39} Shadow deposition is currently the only known way to create combinations of ultra-high vacuum deposited superconductors without involving processing. Alternatives would require either the ability to selectively etch one superconductor without damaging the other or the use of lift-off processes, leading to potential material contamination.

In conclusion, we demonstrate selective shadow deposition to in-plane nanowire networks. The technique is fully based on SAG, making it highly flexible, accurate, reproducible, and scalable. Fully omitting etching of the networks avoids damaging the semiconductor surface, which would lower the device performance. Additionally, it allows for the use of materials to which no selective etchant is currently known. The approach is used for the selective deposition of Al, as well as the first reported growth of epitaxial Pb on InSb nanowires. Shadowing from two sides permits for the combination of multiple materials. This is exemplified in two ways: First, by combining Al as well as Pb with Pt for a complete device fabrication without processing of the nanowire, and second through the combination of Pb with Al for DOS tunnel probing. We believe this approach presents an ideal large-scale platform for the next wave of Majorana based experiments.

\vspace*{2\baselineskip}
{\centering	\textbf{Methods}\par}
\vspace*{1\baselineskip}

\noindent \textbf{Substrate fabrication.}
Fabrication of substrates is a three-step lithography process.\\

\noindent \textit{Markers}: 70~nm Si$_{\text{x}}$N$_{\text{y}}$ layer is deposited by plasma-enhanced chemical vapor deposition (PECVD) on an undoped semi-insulating (111)A InP substrate. AR-P~6200.13 resist is spun at 2000~rpm, baked at 150~°C for 1~min, rinsed in IPA for 30~s, blow dried. Electron-beam lithography (EBL) is used to write marker patterns and developed in AR 600-546 for 1~min. The pattern is transferred into the mask by reactive ion etching (RIE) using CHF$_3$ with added O2. 5~minutes of 50~W~O$_2$ plasma ashing is used to clean the wafer of polymers. Inductively coupled plasma etching (ICP) using CF4/H2 etches the marker pattern ~1.5~$\upmu$m into the substrate. A 3~min buffered hydrofluoric (BHF) etch (NH4F:HF = 7:1) removes the mask.

\noindent \textit{InP}: 20~nm Si$_{\text{x}}$N$_{\text{y}}$ by PECVD. AR-P~6200.13 resist at 4000~rpm, baked at 150~°C for 60~s. The InP wall openings are written by EBL and developed in AR 600-546 for 60~s, rinse in IPA for 30~s, blow dry. RIE, using pure CHF$_3$ RIE to transfer the pattern into the Si$_{\text{x}}$N$_{\text{y}}$ mask. Overnight AR~600-71 stripper to remove the resist, ultrasonic rinse in IPA for 30~s, blow dry. O2 plasma ashing to clean the substrate. A phosphoric acid wet etch (H2O:H3PO4 = 10:1) removes the native substrate oxides, after which the InP walls are grown using a horizontal flow MOVPE reactor. Subsequently the Si$_{\text{x}}$N$_{\text{y}}$ mask is removed by a 3~min BHF etch.

\noindent \textit{InSb}: Analogous to InP processing. 20~nm Si$_{\text{x}}$N$_{\text{y}}$ by PECVD. AR-P~6200.04 at 2000~rpm, baked at 150~°C for 60~s. The InSb network opening are written by EBL and developed in AR~600-546 for 60~s, rinse in IPA for 30~s, blow dry. CHF3 RIE to transfer the pattern. Overnight AR~600-71 stripper to remove the resist, ultrasonic rinse in IPA for 30~s, blow dry. O2 plasma ashing to clean the substrate. A phosphoric acid wet etch \mbox{(H2O:H3PO4 = 10:1)} removes the native substrate oxides, after which the InSb networks are grown using MOVPE.\\

\noindent \textbf{InP homoepitaxy.} Growth takes place in an Aixtron 200 MOVPE horizontal flow reactor with infrared lamp heating. The InP walls are grown at 700~°C using tri-methyl-indium (TMIn) and phosphine~(PH3) with precursor molar fractions $X_{TMIn} = 3.5$ x $10^{-6}$ and $X_{PH3}  = 1.33$ x $10^{-2}$, for 20-50~min depending on the desired wall height. The reactor pressure is 100~mbar, with a total flow of 12000~sccm, using H2 as the carrier gas.\\

\noindent \textbf{InSb heteroepitaxy.} The same reactor is utilized for the InSb growth. An annealing step under phosphine~(PH3) overpressure at 570~°C is used for surface reconstruction of the etched openings and to remove possible oxide residuals from the exposed substrate surface. InSb nanowires were grown at 470 °C using tri-methyl-indium~(TMIn) and tri-methyl-antimony~(TMSb) with precursor molar fractions $X_{TMIn} = 8.4$ x $10^{-8}$ and $X_{TMSb}  =$ 3.7 x $10^{-3}$, for 20-40~min depending on the network size and desired nanowire height. The reactor pressure is 100/50~mbar with a total flow of 12000/6000~sccm respectively and H2 as the carrier gas.\\

\noindent \textbf{Metal deposition.} Nanowire networks are transferred ex-situ to a molecular beam epitaxy (MBE) chamber. An atomic hydrogen clean (20~min under continuous rotation, 400~°C thermocouple, 5 x $10^{-6}$~mbar chamber pressure) is performed to remove the native oxide from the InSb nanowire surface. Subsequently, samples are cooled down to about 110~K by active liquid-nitrogen cooling. Careful alignment of nanowires relative to the metal source is important for well-controlled shadowing of the networks. Al and Pt are deposited at a growth rate of $\sim5.5$~Å~min$^{-1}$, Pb at $\sim24$~Å~min$^{-1}$. Immediately after growth, samples are transferred in-situ to an MBE chamber equipped with an ultrahigh-purity O2 source where they are exposed to $10^{-4}$ mbar of O2 for 15~min. This step is important to form an oxide layer to ‘freeze-in’ the metal film.  This prevents it from diffusing, while the sample warms up to room temperature in ultrahigh vacuum.\\

\noindent \textbf{Focused Ion Beam TEM lamella preparation.} TEM lamellas are prepared in a FEI Nova Nanolab~600i. First, an electron-beam induced C and Pt layer are deposited to minimize the ion beam damage in the following step. After removal of the InP walls using a nanomanipulator, the networks are embedded by an ion beam induced Pt deposition used as a sacrificial layer for the thinning procedure. A TEM lamella is cut and transferred to a half-moon TEM grid for thinning. Finally, thin windows are cut into the lamella using Ga-ion milling in steps at 30~kV, 16~kV and finally 5~kV to create windows with a thickness of less than 100~nm, while minimizing damage induced by the ion beam. It should be noted, that particularly Pb deposited on InSb nanowires suffered from heavy interface intermixing during the thinning procedure. Ion beam milling under cryogenic conditions could possibly solve this.\\

\noindent \textbf{TEM studies.} TEM studies were performed using a probe-corrected JEOL~ARM~20OF, equipped with a 100~mm$^2$ Centurio SDD Energy dispersive X-ray spectroscopy detector. The data processing of the acquired EDX data is described in Figure~S9. Misfit dislocations at the InP-InSb interface were visualized by creating an FFT of an atomic resolution image, applying a filter to only include one set of (111) spots and then creating an inverse FFT.\\

\noindent \textbf{SEM EDX studies.} A FEI Quanta FEG~650 equipped with an Oxford ULTIM MAX 170~mm$^2$ detector was used for the elemental analysis of double-sided shadow deposited samples. An acceleration voltage of 7~kV covered all elements while keeping the probe volume low.


\vspace*{2\baselineskip}
{\centering	\textbf{Acknowledgements}\par}
\vspace*{1\baselineskip}
	We thank P.J.V. Veldhoven for the support with the MOVPE reactor and Martijn Dijstelbloem, Marissa Roijen, Herman Leijssen and Sander Schellingerhout for upkeep of the MBE reactor. This work has been supported by Microsoft Corporation Station-Q and the European Research Council (ERC TOCINA 834290). We acknowledge Solliance, a solar energy R\&D initiative of ECN, TNO, Holst, TU/e, imec and Forschungszentrum Jülich, and the Dutch province of Noord-Brabant for funding the TEM facility.

\vspace*{2\baselineskip}
{\centering	\textbf{Author Contributions}\par}
\vspace*{1\baselineskip}

	J.J. and R.L.M.O.H.V. carried out the fabrication, growth, and growth analysis of the shadowed networks. R.B. supported the Pb deposition. O.V.D.M. aided the device design. C.M. provided the SEM EDX analysis. J.J. made FIB cuts for the TEM analysis. M.A.V. performed the TEM analysis. M.A.V. and E.P.A.M.B. provided key suggestions and discussions and supervised the project. J.J. and R.L.M.O.H.V. wrote the manuscript with contributions from all authors. 

\bibliography{MainText}

\end{document}


\thispagestyle{empty}
\onecolumngrid

\hypersetup{urlcolor=darkblue}
\clearpage
\onecolumngrid
\setlength\parindent{0pt}
\linespread{1.2}
\renewcommand*{\citenumfont}[1]{S#1}
\renewcommand*{\bibnumfmt}[1]{[S#1]}
\renewcommand{\thefigure}{S\arabic{figure}}
\renewcommand{\thetable}{S\arabic{table}}
\renewcommand{\thesection}{S\arabic{section}}
\setcounter{figure}{0}

\begin{center}
	\textbf{\large Supplementary Information: Universal Platform for Scalable Semiconductor-Superconductor Nanowire Networks}
\end{center}
\begin{center}
	\normalsize	{		
		Jason Jung,$^{1,*}$ Roy L.M. Op het Veld,$^{1,*}$ Rik Benoist,$^1$ Orson A.H. van der Molen,$^1$ Carlo Manders,$^2$  Marcel A. Verheijen,$^{1,2}$ Erik P.A.M. Bakkers$^{1,\dagger}$
	}
	\bigskip
	\small
	
	$^*$ These authors contributed equally to this work.
	
	$^{\dagger}$ Correspondence to E.P.A.M.B. (\href{mailto:e.p.a.m.bakkers@tue.nl}{e.p.a.m.bakkers@tue.nl})
	
	\vspace*{0.5cm}
\end{center}	

\vspace*{1.5cm}
\normalsize
\textbf{This document contains:}\\

\textbf{Supplementary Figure~1 $|$} Control over InP wall growth\\
\textbf{Supplementary Figure~2 $|$} Design flexibility\\
\textbf{Supplementary Figure~3 $|$} Scalability and accuracy\\
\textbf{Supplementary Figure~4 $|$} Self-shadowing\\
\textbf{Supplementary Figure~5 $|$} InSb SAG on InP (111)A\\
\textbf{Supplementary Figure~6 $|$} Accuracy of double-sided shadow deposition\\
\textbf{Supplementary Figure~7 $|$} Shadow deposition of Al and Pt\\
\textbf{Supplementary Figure~8 $|$} Shadow deposition of Al and Pb\\
\textbf{Supplementary Figure~9 $|$} EDX data processing\\

\textbf {References}

\clearpage
\begin{figure}
	\centering
	\includegraphics[width=\textwidth]{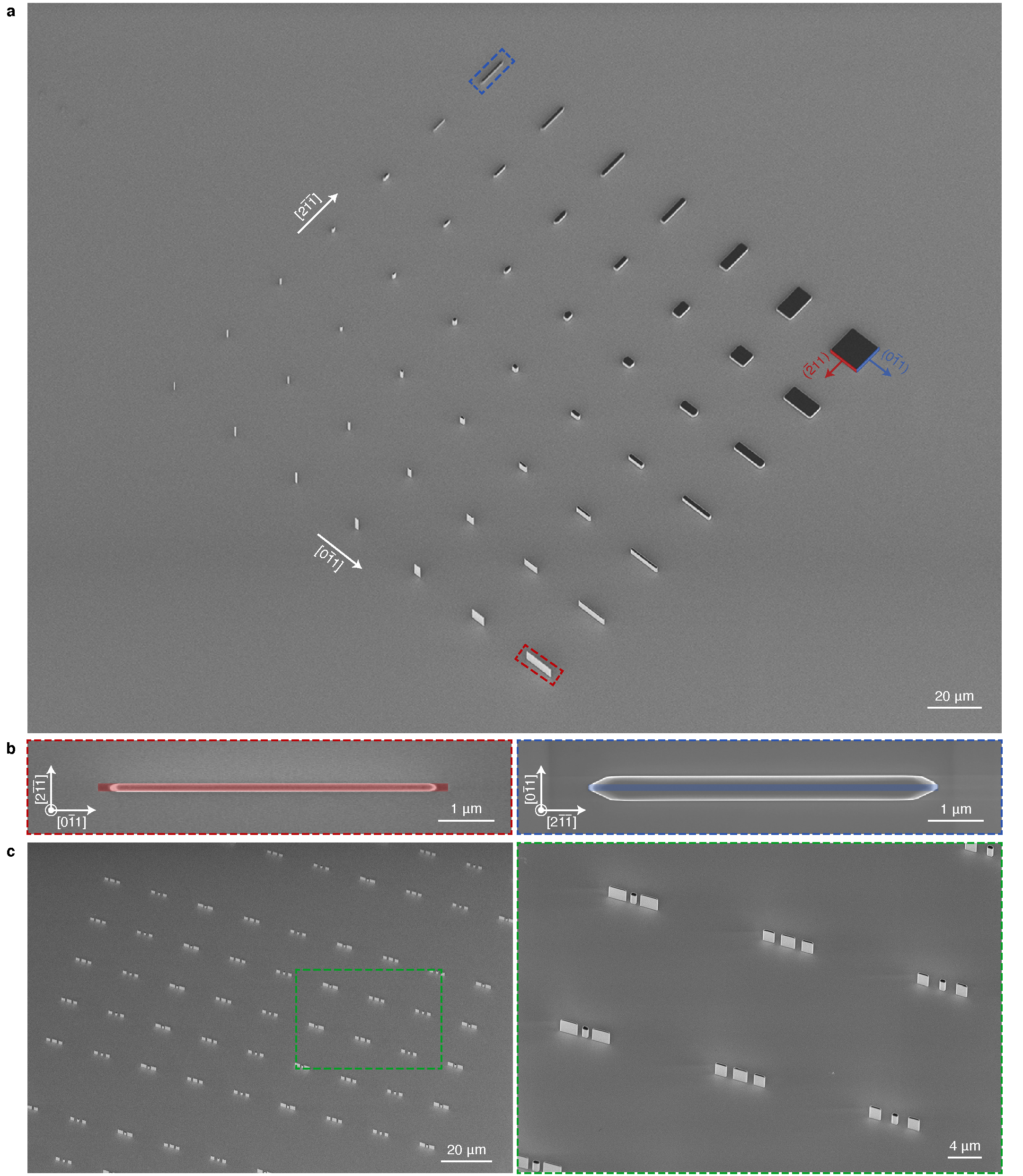}
	\caption{\textbf{Control over InP wall growth.}  \textbf{a,} Tilted SEM micrograph of an array of InP walls. Mask openings patterned along the $\langle0\bar{1}1\rangle$ and $\langle2\bar{1}\bar{1}\rangle$ crystal directions form stable structures bound by $\{0\bar{1}1\}$ and $\{2\bar{1}\bar{1}\}$ side facets. The walls are grown in a variety of lateral dimensions ranging from $100$~nm - 6~$\upmu$m. \textbf{b,} Top-view SEM micrographs of InP walls grown along the $[0\bar{1}1]$ and $[2\bar{1}\bar{1}]$ in-plane direction marked in panel \textbf{a}. It is noted that the former is thinner, reflecting a preferential formation of  $\{2\bar{1}\bar{1}\}$ over $\{0\bar{1}1\}$ facets. This is likely a result of a lower surface energy of the former facet dictated by the V/III precursor ratio and growth temperature\cite{RN728} (see methods). The coloured area indicates the underlying etched mask opening. \textbf{c,} Large array of InP shadow walls with an enlarged image of the area marked in green. Along one axis, the length of the walls is varied, demonstrating the flexibility of the approach. Along the other axis, a copy of the same design was grown indicating the reproducibility of the approach.}
\end{figure}

\clearpage
\begin{figure}
	\centering
	\includegraphics[width=\textwidth]{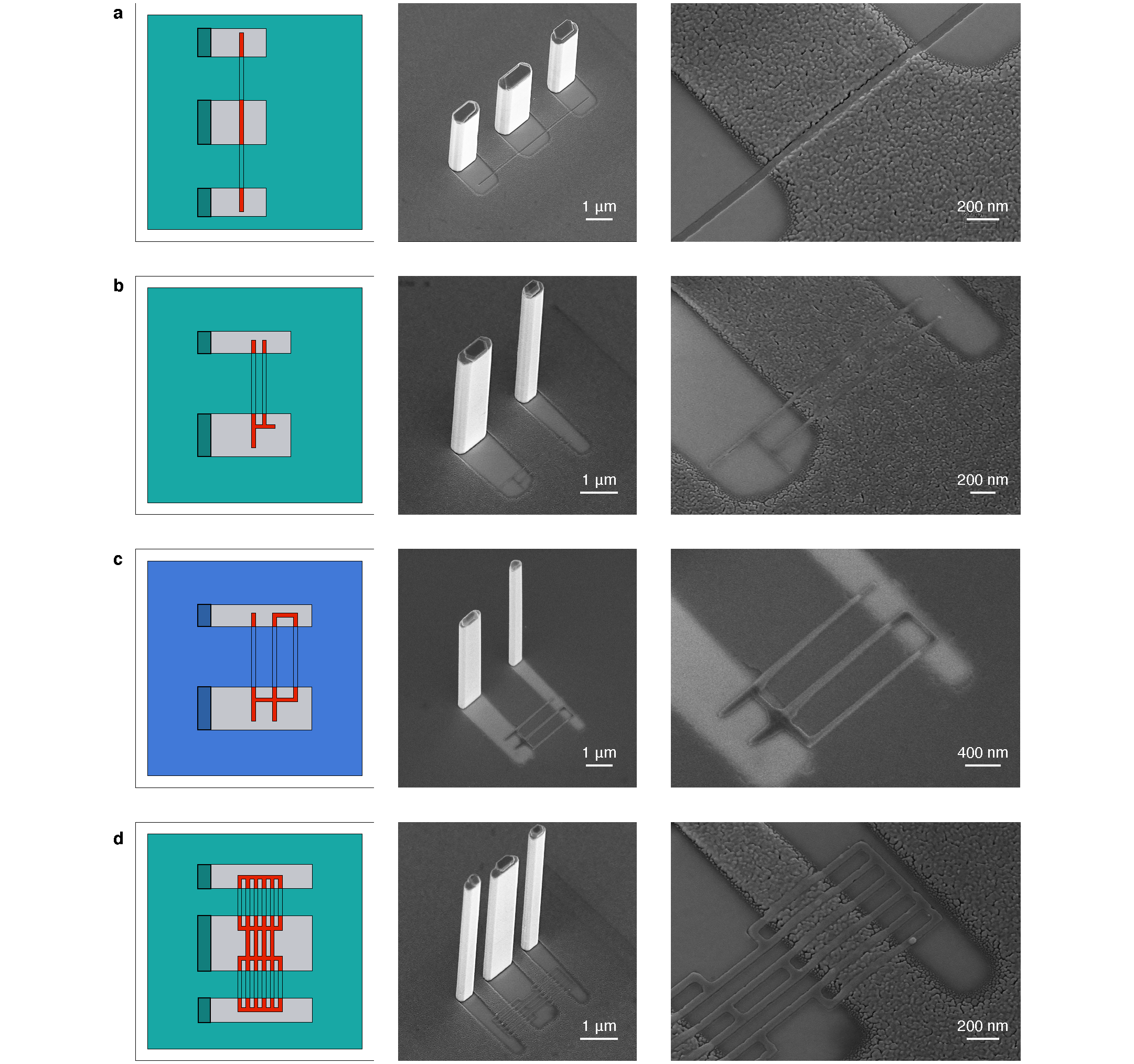}
	\caption{\textbf{Design flexibility.}  \textbf{a,} Single nanowire with double superconducting~(Pb) island design to investigate Majorana fusion rules.\cite{RN812, RN811} The wire is substantially lower than the one shown in the main text, leading to a continuous superconducting film from wire to mask. \textbf{b,} Network based on Majorana teleportation experiments with two superconducting islands~(Pb) on two separate network arms.\cite{RN708} \textbf{c,} Single-qubit nanowire network design with superconducting islands~(Al) on all three arms.\cite{RN708} \textbf{d,} Four-qubit design with superconducting islands~(Pb) on all horizontal arms, demonstrating the scalability of the growth technique.\cite{RN708}}
\end{figure}

\clearpage
\begin{figure}
	\centering
	\includegraphics[width=\textwidth]{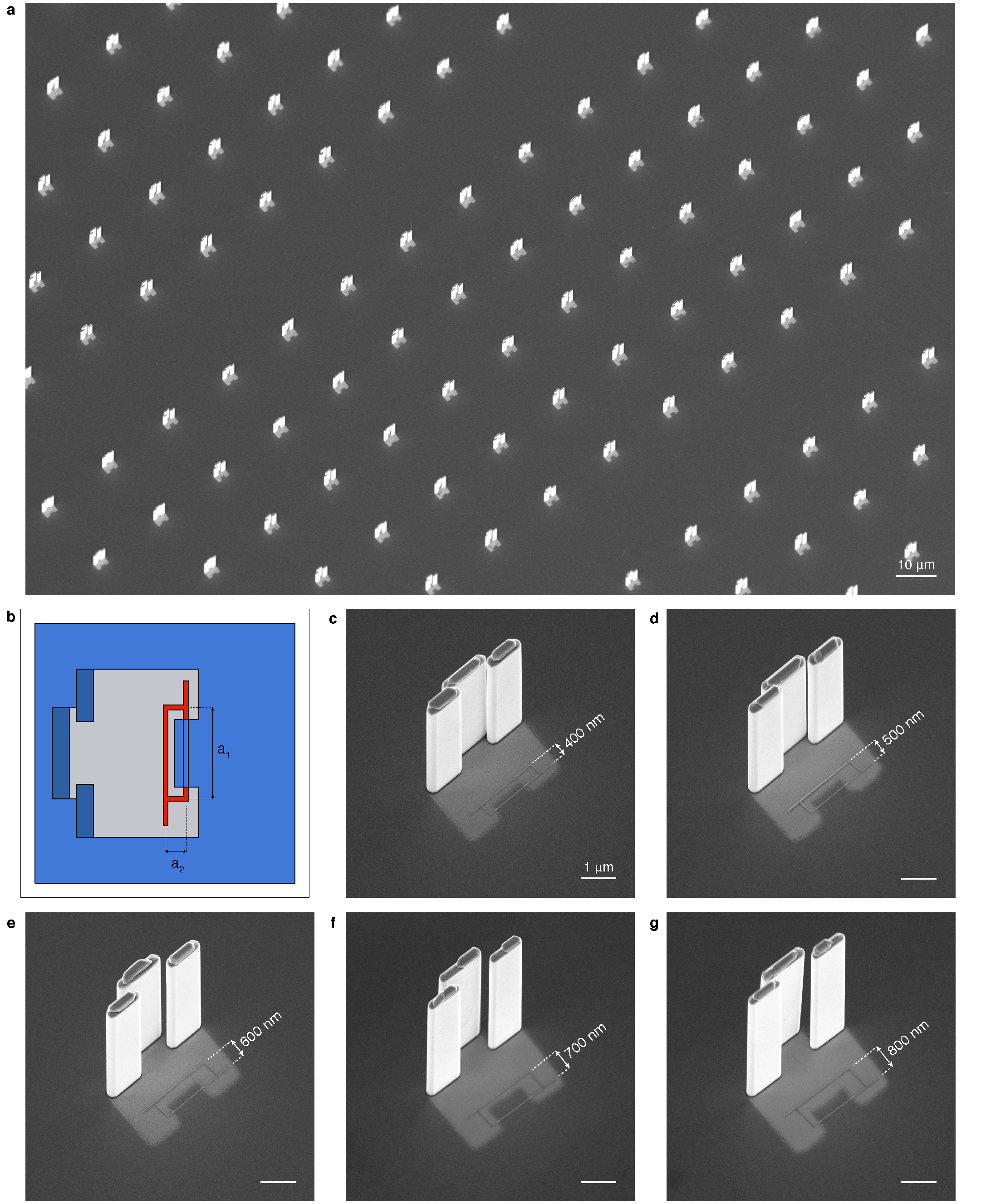}
	\caption{\textbf{Scalability and accuracy.}  \textbf{a,} Tilted SEM micrograph of an array of shadow deposited Aharonov-Bohm interferometer. \textbf{b,} Illustration of the design indicating the length of the two arms $a_1$ and $a_2$. \textbf{c-g,} A series of devices taken from the sample shown in panel \textbf{a}. The length $a_1$ is kept constant at 2~$\upmu$m, while varying $a_2$ from 400 - 800~nm. The network shown in panel \textbf{c} has the same design as Figure 1h but is not the same device.}
\end{figure}

\clearpage
\begin{figure}
	\centering
	\includegraphics[width=\textwidth]{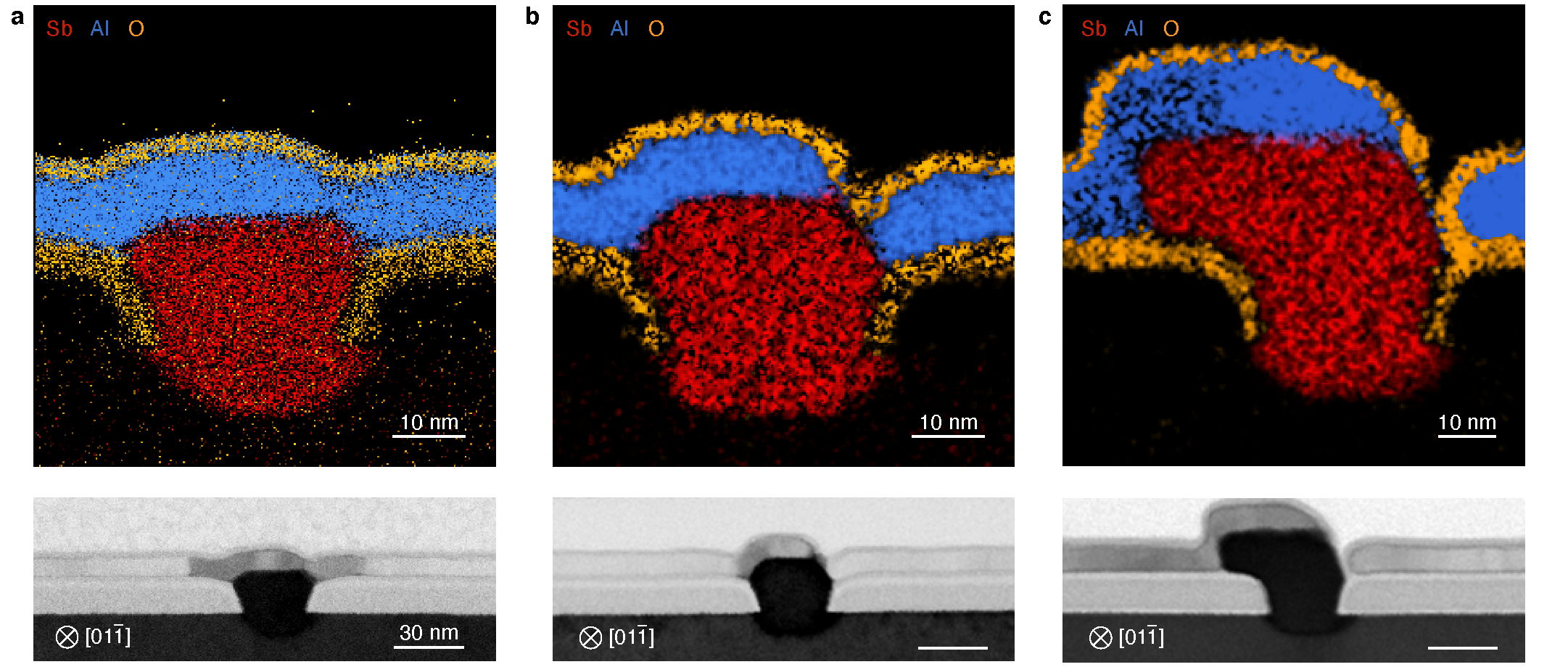}
	\caption{\textbf{Self-shadowing.}  \textbf{a,} EDX elemental mapping of an InSb nanowire (red) cross-section with a superconducting Al film~(blue) deposited on top. Here, the nanowire is grown to approximately the height of the mask, leading to a continuous film to the mask on both sides of the nanowire. No oxide~(orange) interruptions are present in the Al layer. The HAADF scanning TEM micrograph below shows the continuous film. \textbf{b,} The nanowire is grown slightly higher than the mask, allowing for self-shadowing due to the superconductor deposition at $\alpha = 25\degree$. Two separate Al segments are created, both connected to the wire with an oxide-free interface. \textbf{c,} Increasing the nanowire height fully interrupts the deposited Al film. At this height, the right side of the film is completely disconnected from the rest. A thin oxide layer has formed electrically isolating this section of the Al film from the nanowire.}
\end{figure}

\clearpage
\begin{figure}
	\centering
	\includegraphics[width=\textwidth]{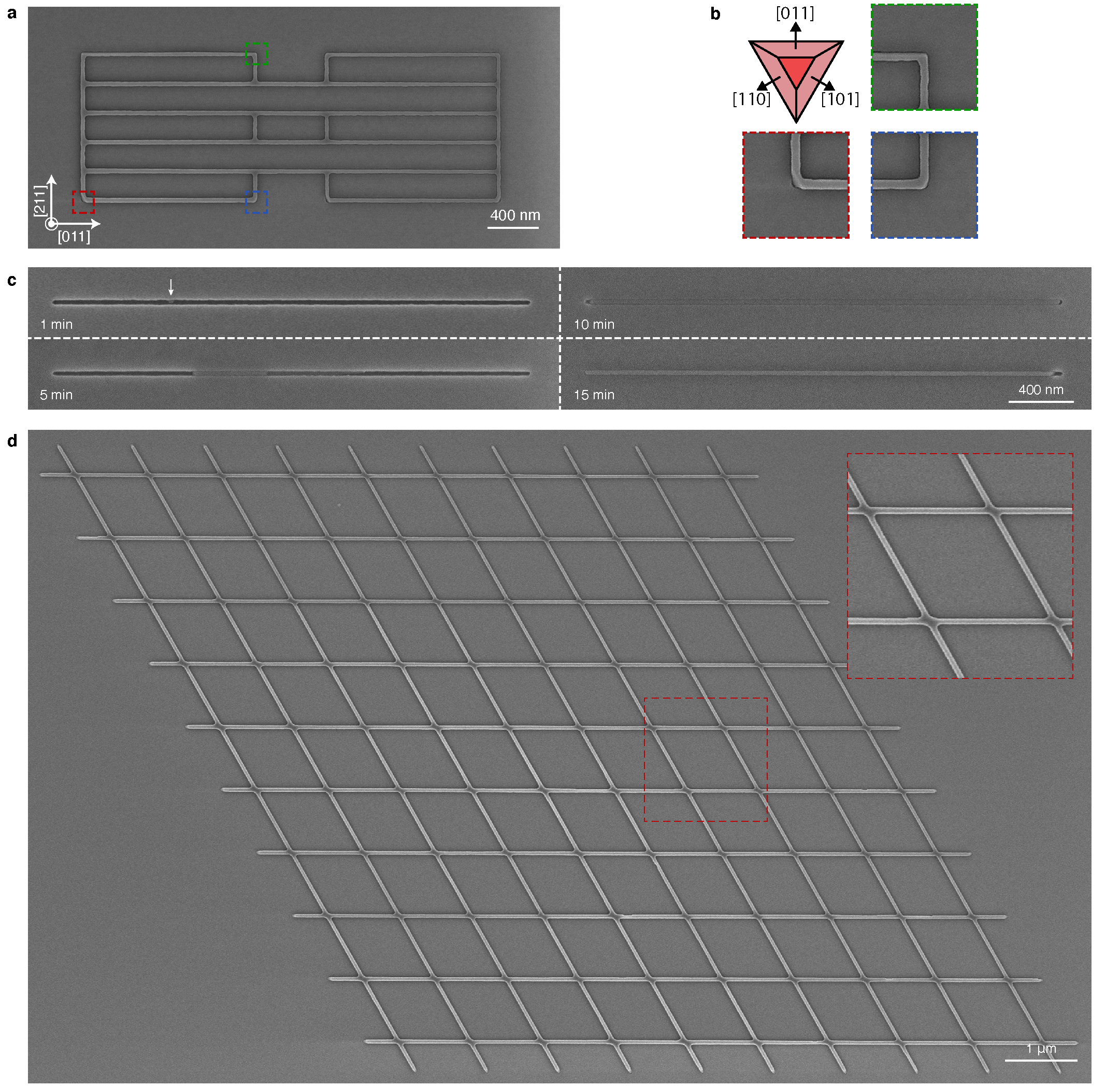}
	\caption{\textbf{InSb SAG on InP (111)A.}  \textbf{a,} SEM micrograph of a four-qubit design InSb nanowire network grown on an InP~(111)A substrate using a Si$_{\text{x}}$N$_{\text{y}}$ selective area mask. \textbf{b,} Terminating $\{110\}$ growth facets are visible at the corners of the design. \textbf{c,} Growth as a function of time. A single nucleation site (arrow) has formed inside the mask opening at 1 min and expands over time. After 10 min the nanowire is fully grown and exceeds the mask at 15~min. \textbf{d,} Large scale network indicating the scalability of InSb SAG on InP~(111)A.}
\end{figure}

\clearpage
\begin{figure}
	\centering
	\includegraphics[width=\textwidth]{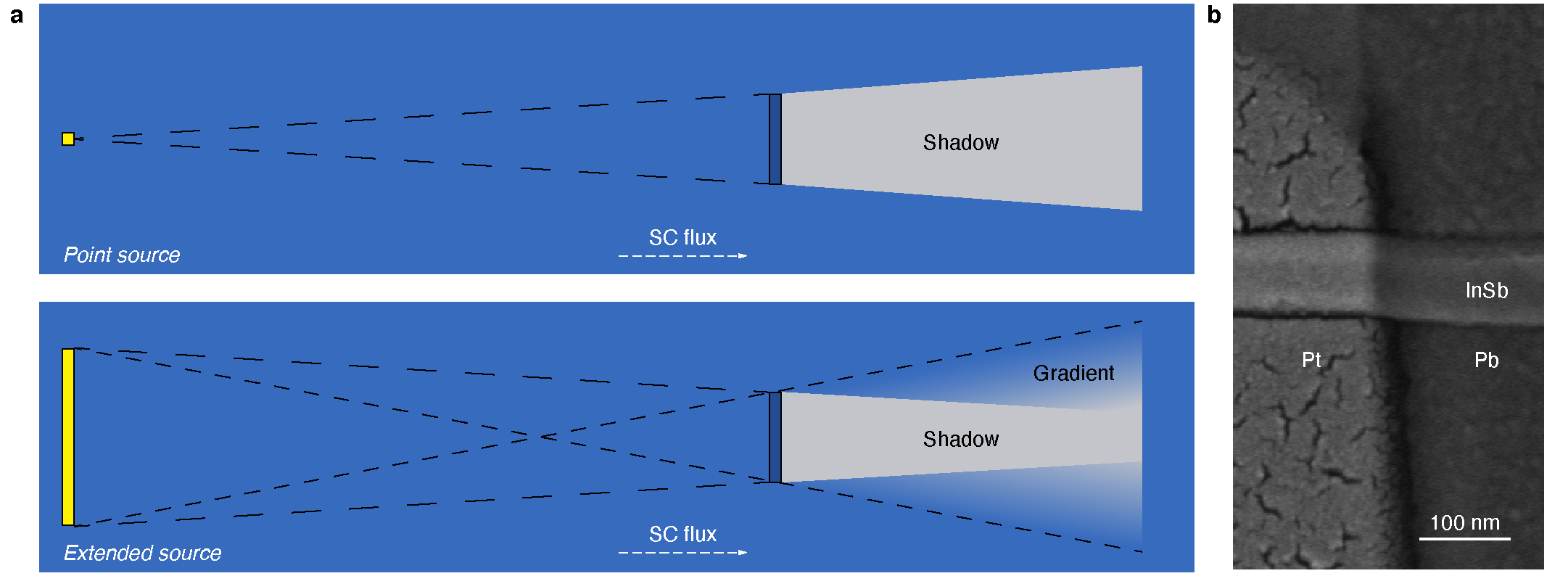}
	\caption{\textbf{Accuracy of double-sided shadow deposition.}  \textbf{a,} An idealized point-like metal source creates a sharp shadow. In reality, metal sources have a finite spatial extension. A result of this is a gradient in which the thickness of the shadowed metal increases. The extension of this gradient increases with the distance to the wall. Other factors possibly limiting the accuracy of the shadow deposition are overgrowth of the walls (compare Figure S1), processing related imprecisions (e.g., e-beam lithography, widening of the mask openings during the dry etching), and the angle alignment of the sample for the metal deposition. \textbf{b,}~Close-up SEM of an in-plane grown InSb nanowire with a normal metal~(Pt) selectively deposited on the left side and a superconductor~(Pb) on the right. Despite the aforementioned challenges a $\sim15$~nm shadow was achieved.}
\end{figure}
\clearpage

\twocolumngrid

\clearpage
\begin{figure}
	\centering
	\includegraphics[width=.48\textwidth]{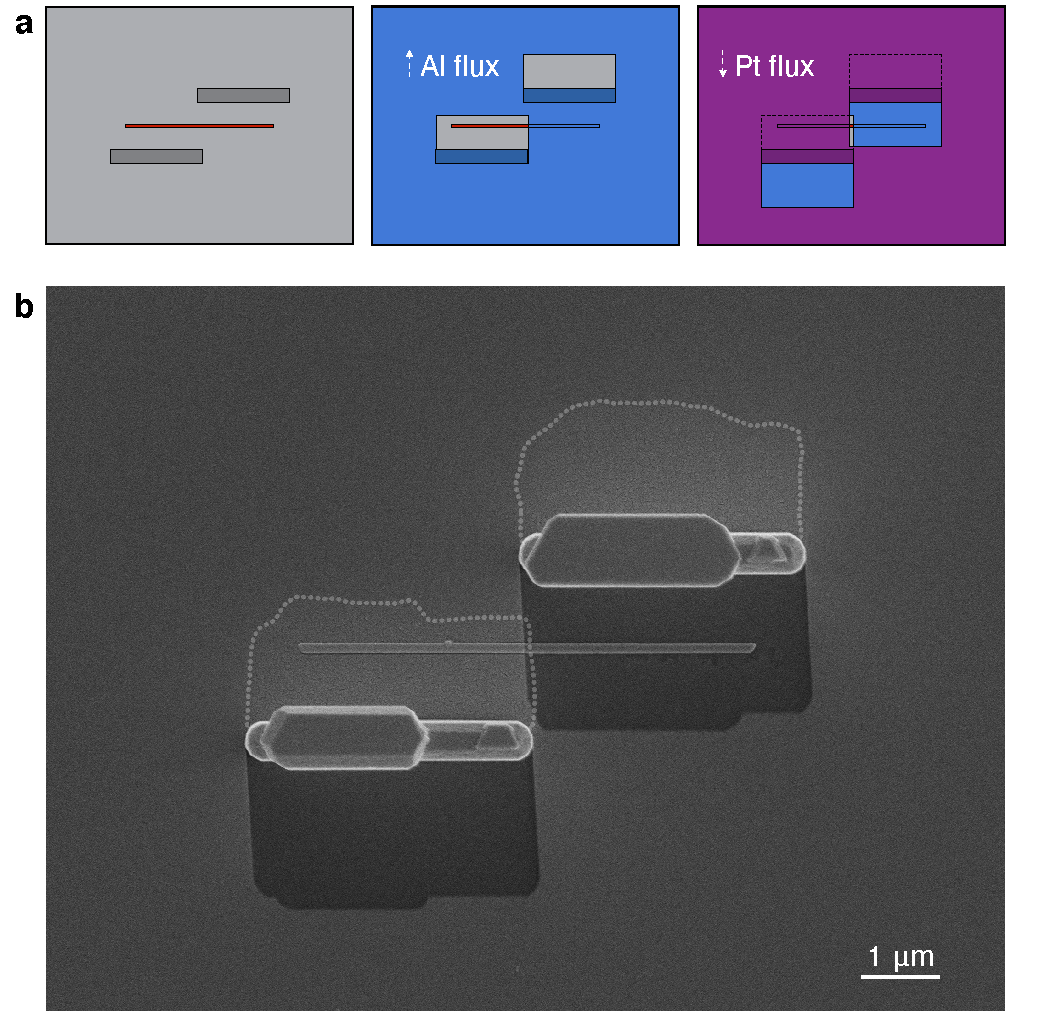}
	\caption{\textbf{Shadow deposition of Al and Pt.}  \textbf{a,}~Schematic top-view of two-sided metal deposition on an InSb nanowire~(red) using InP walls. The aluminium (blue) and Pt (purple) are deposited from opposite sides, creating a normal metal-semiconductor-superconductor junction. \textbf{b,} Top-view SEM micrograph of such a device.}
\end{figure}

\clearpage
\begin{figure}
	\centering
	\includegraphics[width=.48\textwidth]{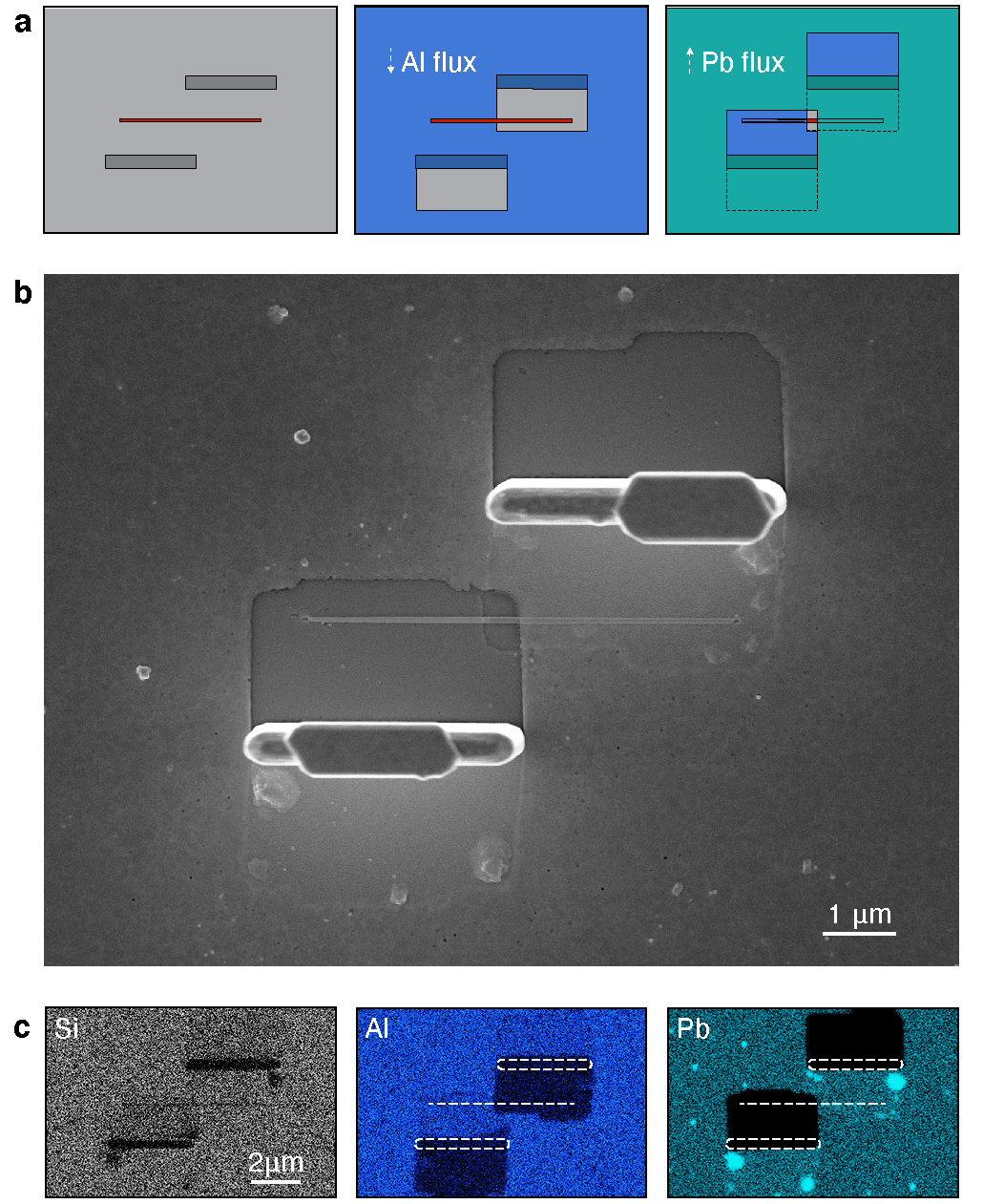}
	\caption{\textbf{Shadow deposition of Al and Pb.}  \textbf{a,}~Schematic top view of two-sided superconductor deposition on an InSb nanowire. Al~(blue) and Pb~(cyan) are deposited from opposite directions, creating two separated superconducting segments on the nanowire. \textbf{b,} Top-view SEM micrograph of such a device. \textbf{c,} SEM-EDX elemental mapping of the device shown in \textbf{b}. The silicon signal shows the position of the walls and nanowire in the mask. The Al and Pb signals show the shadowed region. Some Pb accumulation can be observed on the mask.}
\end{figure}
\clearpage

\onecolumngrid
\clearpage
\begin{figure}
	\centering
	\includegraphics[width=\textwidth]{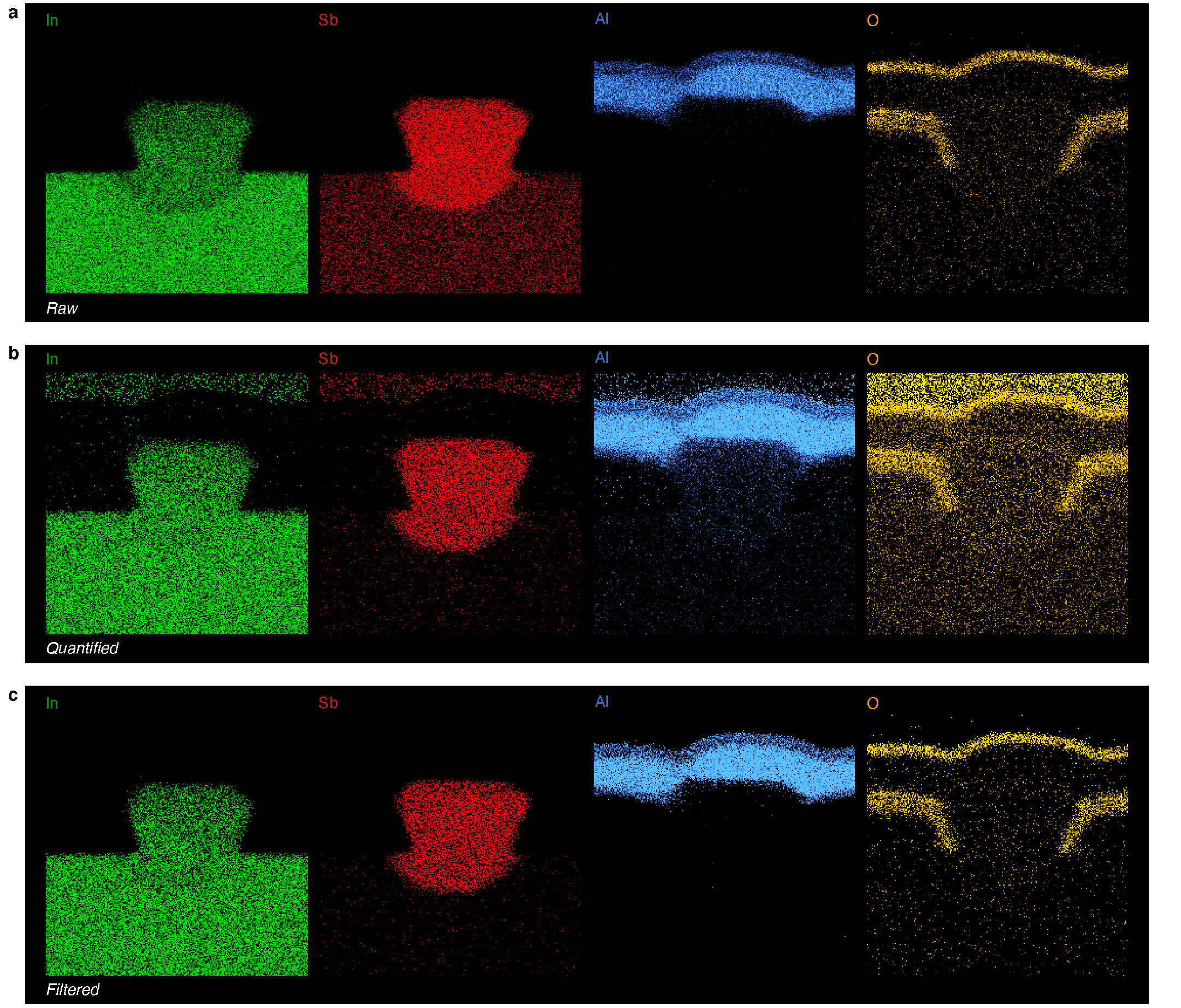}
	\caption{\textbf{EDX data processing.}  \textbf{a,} Raw data of an EDX elemental mapping. Two imaging artefacts are visible in the raw counts maps that support the choice for using quantified data for representation. Firstly, the In signal (green) has different intensities in the nanowire and the substrate due differences in In-L X-ray yield from InP and InSb regions in the sample. Secondly, the Sb signal (red) shows counts inside the substrate, due to a peak overlap between the Sb-L and In-L lines. \textbf{b,} Both of these issues can be accounted for by quantifying the data, which normalizes the number of counts for each element and separates contributions of overlapping peaks. It however also significantly increases noise. This is most prominent in areas with a net low X-ray yield, such as the protective carbon layer covering the AlO$_\text{x}$. \textbf{c,} Masking \textbf{b} with \textbf{a}, shows the quantified data only where there is X-ray yield in the raw data and reduces the introduced background noise. EDX data is shown in this way throughout the paper.}
\end{figure}

\clearpage
\bibliographystyle{naturemag}
\bibliography{Supplementary.bib}